\documentclass[aps,pre,twocolumn,superscriptaddress,10pt]{revtex4-2}
\usepackage{array}[=2016-10-06]
\usepackage[utf8]{inputenc}
\usepackage{color}
\usepackage{amsmath}
\usepackage{amssymb}
\usepackage{xcolor}
\usepackage{graphicx}
\usepackage{esint}
\usepackage{appendix}
\usepackage{comment}
\usepackage{CJK}
\usepackage{hyperref}
\usepackage{diagbox}
\usepackage{bm}
\usepackage{array}
\usepackage{tabularx}
\graphicspath{{./}}

\makeatletter
\@ifundefined{textcolor}{}
{%
 \definecolor{BLACK}{gray}{0}
 \definecolor{WHITE}{gray}{1}
 \definecolor{RED}{rgb}{1,0,0}
 \definecolor{GREEN}{rgb}{0,1,0}
 \definecolor{BLUE}{rgb}{0,0,1}
 \definecolor{CYAN}{cmyk}{1,0,0,0}
 \definecolor{MAGENTA}{cmyk}{0,1,0,0}
 \definecolor{YELLOW}{cmyk}{0,0,1,0}
}
\makeatother

\begin{document}
\begin{CJK*}{UTF8}{gbsn}
\title{Exact First-Passage Time Response Theory from Steady-State Response}%Mean First-Passage Time Response Relations}

\author{Ruicheng Bao}
\email{Corresponding author: ruicheng@g.ecc.u-tokyo.ac.jp}
\affiliation{Department of Physics, Graduate School of Science,
The University of Tokyo, Hongo, Bunkyo-ku, Tokyo 113-0033, Japan}

\author{Shiling Liang (梁师翎)}
\email{Corresponding author: shiling@pks.mpg.de}
\affiliation{Center for Systems Biology Dresden, 01307 Dresden, Germany}
\affiliation{Max Planck Institute for the Physics of Complex Systems, 01187 Dresden, Germany}
\affiliation{Max Planck Institute of Molecular Cell Biology and Genetics, 01307 Dresden, Germany}

\begin{abstract}
The mean first-passage time (MFPT) provides a universal temporal measure of transport, reaction,
search, and switching processes in physical, chemical, and biological systems. Understanding how
MFPTs respond to perturbations is therefore crucial for prediction and control, yet a systematic theory has been lacking. We establish a compact theoretical framework for linear and nonlinear MFPT response in continuous-time Markov processes. The key tool is an exact correspondence that maps the intrinsically transient response of MFPTs onto the steady-state response of an auxiliary system. This correspondence yields exact and universal response relations for MFPTs between arbitrary state pairs, expressed entirely in terms of unperturbed MFPTs and steady-state probabilities.
%The response of the steady-state distribution to arbitrary perturbation is also obtained with these unperturbed quantities. 
%Using the exact relation, 
We then obtain a factorized physical decomposition of the MFPT response into linear upstream, linear downstream, and nonlinear contributions. Further corollaries include response-curve inference rules, fundamental bounds on MFPT responses, analytical expressions for higher-order responses of MFPTs and steady-state probabilities, and multi-rate response formulas. Additionally, our result offers computational advantages in calculating both MFPTs and steady-state distributions. Finally, %as applications, 
a biologically motivated %bottleneck/rescue 
folding network is analyzed, and a recently reported paradox on MFPT is clarified.
\end{abstract}
\maketitle
\end{CJK*}

\textit{Introduction---}The mean first-passage time (MFPT) is the average time for a stochastic process to reach a prescribed target state for the first time. This fundamental concept underpins the characterization of transport processes across a wide array of disciplines \cite{redner2001guide,benichou2007first,benichou2016mean}. Its significance spans from physics, biology and sociology to computer science \cite{04prlmfpt}, offering profound insights into diverse phenomena. For instance, it quantifies the mean transition time between the clockwise and counterclockwise states of molecular motors \cite{00science_ultra}, the mean dwell time between protein folding and unfolding \cite{yu2023unidirectional}, transcriptional burst cycle time in gene regulation \cite{11scienceburst,19natureburst,24PNASlammer}, or the average time it takes for a virus to successfully infect a cell \cite{holcman2007modeling}. Beyond these molecular and cellular processes, MFPT governs the search strategies of animals for food \cite{benichou05search}, dictates the completion time of complex biochemical reaction networks \cite{benichou2007first,benichou2008enhanced}, measures ecological resilience \cite{21Scienceecology} and determines the efficiency of computational algorithms \cite{24fptalgorithm}, establishing itself as a cornerstone for understanding dynamic processes in both natural and artificial systems.

Understanding how systems respond to external perturbations represents one of the most fundamental approaches in natural science. Given the great importance of MFPT, quantitatively understanding how this crucial temporal metric responds to external perturbations becomes essential for both theoretical development and practical applications. Such response behavior is particularly relevant in biological systems, where cellular processes must maintain robustness against environmental fluctuations, and in engineered systems requiring precise temporal control. While the fluctuation-dissipation theorem provides a powerful framework for equilibrium systems \cite{kubo1966Fluctuationdissipation}, its extension to nonequilibrium scenarios remains challenging, with recent advances primarily focused on steady-state quantities \cite{meyerjr.1980Condition,cho2000Markov,li2003Sensitivity,lan2012Energy,govern2014Energy,owen2020Universal,gao2022thermodynamic,chun2023Tradeoffs,owen2023Size,fernandesmartins2023Topologically,owen2023Thermodynamic,harvey2023Universal,aslyamov2024Nonequilibrium,aslyamov2024General,khodabandehlou2024Affine,kwon2024Fluctuationresponse,mahdavi2024flexibility,ptaszynski2024Critical,ptaszynski2024Dissipation,ptaszynski2024Nonequilibrium,bao2024nonequilibrium,cao2025Stochastic}. Although several studies have examined the response of non-stationary observables \cite{maes09response,dechant2020Fluctuation,maes2020Response,harunari2024Mutual,dechant2024Fundamental,zheng2024Universal}, their findings cannot be directly applied to MFPT, as they focus on fixed time intervals rather than processes involving random stopping times. A comprehensive theoretical framework for characterizing and understanding the physics of MFPT response remains elusive.

Here, we fill this critical gap by deriving exact relations that govern the response of MFPTs between arbitrary states in continuous-time Markov processes. We achieve this by establishing a conceptually new tool: an equivalence between MFPT response in the original system and steady-state response in an auxiliary fast-reset system. This equivalence translates steady-state response identities into exact MFPT-response identities, thereby yielding a comprehensive response framework. Our framework reveals that both linear and nonlinear responses of MFPTs can be precisely determined using a limited set of unperturbed MFPTs and steady-state probabilities, marking a significant advance in our ability to predict and control temporal dynamics in nonequilibrium systems. This stands in sharp contrast to the only directly related prior work we are aware of, a study on discrete-time systems \cite{hunter05sensitivity}, where the MFPT-sensitivity formulae are either not closed or involve quantities without clear physical meaning.  Remarkably, despite MFPT being a global observable, the number of required unperturbed quantities to express the perturbed MFPT depends solely on the number of perturbed transition rates, independent of system size and topology. Physically, we show that a single-rate MFPT response factorizes into source-side accessibility, target-side gain/loss, and target-side nonlinear screening. This structure gives a physically transparent bottleneck criterion, explains how finite response is screened, and turns MFPT measurements into a diagnostic tool for inferring the dynamical importance of a perturbed transition.

\textit{Setup and steady-state response relation---}We consider the continuous-time Markov dynamics described by the master equation
\begin{equation*}
    \frac{d\boldsymbol{p}(t)}{dt}=W\boldsymbol{p}(t),
\end{equation*}
which provides a fundamental framework across diverse mesoscopic systems, such as biochemical systems from cellular processes to chemical reactions. It also plays a crucial role in communication and information theory \cite{shannon1948mathematical}. Additionally, the Markov process represents a first-order approximation to general dynamical equations \cite{mahdavi2024flexibility}. Here, $\boldsymbol{p}(t)$ is the probability distribution at time $t$ and $W$ is the transition rate matrix, whose off-diagonal entry $W_{ij}\geq 0$ ($i\neq j$) denotes the transition rate from state $j$ to state $i$  and whose diagonal entries are defined as $W_{ii}=-\sum_{j\neq i}W_{ji}$. Our only assumption is that the Markov chain is irreducible, with a steady-state distribution $\boldsymbol{\pi}$ satisfying $\sum_i\pi_i=1$ and $\pi_i\geq 0$. We define $\tau_{kl}$ as the MFPT from state $l$ to state $k$, with $\tau_{ii}=0$ for any $i$.

If a transition rate $W_{mn}$ of the original dynamics is perturbed to $W_{mn}\to W_{mn}+\Delta W_{mn}$, with $\Delta W_{mn}\geq -W_{mn}$, then the steady-state $\pi_k^{\prime}$ of the perturbed system is given by the identity $\pi_k^{\prime}-\pi_k=(\tau_{kn}-\tau_{km})\Delta W_{mn}\pi_n^{\prime}\pi_k$ \cite{bao2024nonequilibrium}.  After solving the $k=n$ case, the closed-form expression for any $\pi'_k$ follows as:

\textit{Lemma 1.}

\begin{equation} \label{lemma1}
    \pi^{\prime}_{k}=\pi_{k}\left[1+\frac{(\tau_{kn}-\tau_{km})\Delta W_{mn}\pi_{n}}
    {1+\Delta W_{mn}\pi_{n}\tau_{nm}}\right].
\end{equation}
This establishes how a steady-state distribution responds to an arbitrarily strong perturbation on a single edge.

\textit{MFPT and steady-state response correspondence---}We first propose a novel observation, namely, the equivalence between the MFPT response and the steady-state response:
\begin{equation}\label{keyidea}
    \frac{\tau_{kl}}{\tau_{kl}^{\prime}}=\lim_{W_{lk}^{*}\to\infty}\frac{\pi_k^{*\prime}}{\pi_k^{*}}.
\end{equation}
Here, $\pi_k^{*}$ is the steady-state probability of state $k$
in an auxiliary system obtained by adding a transition $k\to l$
with rate $W^{*}_{lk}$ to the original dynamics. The primed quantities
refer to the same perturbation $W_{mn}\to W_{mn}+\Delta W_{mn}$ applied
to the original system and to this auxiliary system, respectively:
$\tau_{kl}^{\prime}$ is the perturbed MFPT, while $\pi_k^{*\prime}$ is
the perturbed auxiliary steady-state probability. 
The observation is unexpected, given that it connects the responses of two seemingly unrelated (and even opposite) physical quantities. One reflects the stationary properties of the system, while the other characterizes the non-stationary properties. A physically motivated derivation is given in End Matter: a fast reset from the target $k$ to the initial state $l$ ($W^{*}_{lk}\to \infty$) turns repeated first-passage events into stationary cycles, so the MFPT becomes the inverse stationary flux through the reset edge. The linear-response version of Eq.~\eqref{keyidea} is $\partial_{W_{mn}}\ln \tau_{kl}=-\lim_{W_{lk}^{*}\to\infty}\partial_{W_{mn}}\ln\pi_k^*$. Eq.~\eqref{keyidea} is the key bridge that turns a steady-state response identity into an exact MFPT response theory.

\textit{Linear and nonlinear response identities---}With this key observation, we can determine the exact response of MFPTs using the response formula for steady-state probability (Lemma~\ref{lemma1}). The central result is that, for a single perturbed transition rate $W_{mn}$,
\begin{subequations}\label{main_result}
\begin{align}
    \partial_{W_{mn}}\tau_{kl}
    &=-\pi_n(\tau_{kn}-\tau_{km})(\tau_{nk}+\tau_{kl}-\tau_{nl}),
    \label{mfptlinear}\\
    \frac{\tau_{kl}^{\prime}-\tau_{kl}}{\Delta W_{mn}}
    &=\frac{\partial_{W_{mn}}\tau_{kl}}
    {1+\Delta W_{mn}\pi_n(\tau_{kn}+\tau_{nm}-\tau_{km})}.
    \label{nonlinearlinear}
\end{align}
\end{subequations}
Eq.~\eqref{mfptlinear} gives the exact linear response, while Eq.~\eqref{nonlinearlinear} gives the exact relation between nonlinear and linear response for an arbitrarily strong single-rate perturbation. The proof is sketched in End Matter, while the full details are given in the Supplemental Material (SM) Sec. II \cite{supplemental_material}.

To expose the physical content of Eq.~\eqref{nonlinearlinear}, we introduce (The positivity is due to the triangle inequality $\tau_{km} +\tau_{mn} \geq \tau_{kn}$ for MFPTs \cite{supplemental_material,bao2024nonequilibrium})
\begin{align*}
    U_{k\leftarrow l|n}&:=\tau_{nk}+\tau_{kl}-\tau_{nl}\geq0,\\
    G_{k|mn}&:=\tau_{kn}-\tau_{km},\\
    \Sigma_{k|mn}&:=\tau_{kn}+\tau_{nm}-\tau_{km}\geq 0,
\end{align*}
so that the nonlinear response can be expressed as
\begin{equation}\label{nonlinear_identity}
    \tau_{kl}^{\prime}-\tau_{kl}
    =-\frac{\pi_n\Delta W_{mn}\,G_{k|mn}\,U_{k\leftarrow l|n}}
    {1+\pi_n\Delta W_{mn}\Sigma_{k|mn}},
\end{equation}
whose transparent physical interpretation is as follows.

\textit{(i) Upstream accessibility.} The factor $U_{k\leftarrow l|n}\geq 0$ measures how strongly first-passage trajectories from $l$ to $k$ go through the perturbed source state $n$ before reaching the target. If $U_{k\leftarrow l|n}=0$, the perturbation is not sampled on the way to $k$, and the response vanishes. It is therefore a source-side bottleneck factor: when $U_{k\leftarrow l|n}$ is large, trajectories are strongly funneled through $n$, so the perturbation is sampled often and the linear response is correspondingly amplified. %This is one ingredient for approaching the universal upper envelope of the relative response.}

\textit{(ii) Downstream gain or loss.} The factor $G_{k|mn}$ decides the sign of the response and is independent of the initial state. If $\tau_{km}<\tau_{kn}$, then state $m$ is closer to the target than state $n$, and increasing $W_{mn}$ accelerates the first-passage process, so $\partial_{W_{mn}}\tau_{kl}<0$. If $\tau_{km}>\tau_{kn}$, the perturbation drives trajectories to a state farther from the target, so the MFPT increases. The larger the asymmetry $|\tau_{kn}-\tau_{km}|$, the more kinetically distinct the two endpoints are relative to the target, and the larger the response amplitude. When $\tau_{km}=\tau_{kn}$, the two endpoints are MFPT-symmetric with respect to $k$, and the MFPT response is exactly zero for every initial state $l$. %and for every perturbation strength.

\textit{(iii) Downstream nonlinear screening.} The factor $\Sigma_{k|mn}\geq 0$ is independent of the initial state, $l$ and is a purely finite-perturbation contribution, which vanishes in the linear-response limit. %It therefore isolates the purely target-side renormalization of finite response.
When $\Sigma_{k|mn}=0$, the perturbed edge is a bottleneck: trajectories that enter $m$ must effectively return through $n$ before reaching $k$, so no nonlinear screening remains and Eq.~\eqref{nonlinearlinear} reduces to the tangent extrapolation $\tau_{kl}^{\prime}-\tau_{kl}=\Delta W_{mn}\partial_{W_{mn}}\tau_{kl}$. When $\Sigma_{k|mn}$ is large, even a small perturbation can generate visible nonlinear corrections and invalidate the linear-response prediction. Physically, a large $\Sigma_{k|mn}$ reflects downstream bypasses that strongly suppress the finite response. Equation~\eqref{nonlinearlinear} then has an inverse utility: by measuring the linear response and one nonlinear MFPT change, one can infer the bottleneck size $\Sigma_{k|mn}$ and decide whether the perturbed edge is dynamically important. %without knowing the microscopic details of the perturbed edge.}

%A direct corollary of Eq.~\eqref{mfptlinear} is that, for the physical parametrization \cite{dechant2020Fluctuation, owen2020Universal}
%$W_{mn}=e^{-A_{mn}}=e^{-B_{mn}+E_n}$ ($B_{mn}=B_{nm}$ represents the barrier perturbation, while $E_n$ is the energy perturbation), the same linear identity yields the summation theorems
% \begin{equation}\label{sumrules}
%     \sum_m \partial_{E_m}\ln\tau_{kl}=-1,
%     \qquad
%     \sum_{m<n}\partial_{B_{mn}}\ln\tau_{kl}=1.
% \end{equation}
% These are the MFPT counterparts of summation identities in control analysis, and we collect the derivation in \cite{supplemental_material}. 

Equation~\eqref{nonlinearlinear} further implies that $\tau_{kl}$ is \textit{monotonic} in any perturbed rate $W_{mn}$, and that it is \textit{concave} for harmful edges (increasing its weight decelerates the first-passage event) but \textit{convex} for helpful edges (increasing its weight accelerates the first-passage event), as shown in Sec. V of SM \cite{supplemental_material}.

\textit{Full response curve from two MFPT measurements---}
Eq.~\eqref{nonlinearlinear} implies that the entire response curve can be reconstructed from the unperturbed MFPT $\tau_{kl}$, its linear response $\partial_{W_{mn}}\tau_{kl}$, and one additional finite measurement. Denoting that reference nonlinear point by $\tau_{kl}^{\mathrm{ref}}:=\tau_{kl}|_{\Delta W=\Delta W_*}$, we obtain
\begin{align}
\tau_{kl}(\Delta W)
&=\tau_{kl}+\frac{\Delta W\,\partial_{W_{mn}}\tau_{kl}}
{1+\Delta W\,\pi_n\Sigma_{k|mn}},
\label{fullcurve}\\
\pi_n\Sigma_{k|mn}
&=\frac{\partial_{W_{mn}}\tau_{kl}}{\tau_{kl}^{\mathrm{ref}}-\tau_{kl}}-\frac{1}{\Delta W_*}.\nonumber
\end{align}
Thus, a single nonlinear data point $\tau_{kl}^{\mathrm{ref}}$ turns the linear response $\partial_{W_{mn}}\tau_{kl}$ of the original system into a global finite-amplitude prediction $\tau_{kl}(\Delta W)$ for any $\Delta W$.

\textit{Fundamental bounds on the MFPT response---}By the non-negativity of $\tau^{\prime}_{kl}$, Eq.~\eqref{nonlinearlinear} yields two new inequalities among MFPTs in the unperturbed network (see Sec. VI B of \cite{supplemental_material}):
\begin{subequations}
\begin{align}
&(\tau_{kn}-\tau_{km})(\tau_{nk}-\tau_{nl})
\le \tau_{kl}\tau_{nm},
\label{mfptineq1}\\
&\phi_{mn}\left(\tau_{nm}-\frac{(\tau_{kn}-\tau_{km})(\tau_{nk}-\tau_{nl})}{\tau_{kl}}\right)
\le 1,
\label{mfptineq2}
\end{align}
\end{subequations}
which may be of independent interest.  Here, $\phi_{mn}:=W_{mn}\pi_n$ is the directed traffic. They follow from  two extreme perturbation limits: $\lim_{\Delta W_{mn}\to\infty}\tau^{\prime}_{kl}\geq 0$ gives
Eq.~
\eqref{mfptineq1}, whereas $\tau^{\prime}_{kl}|_{\Delta W_{mn}=-W_{mn}}\geq 0$ gives Eq.~\eqref{mfptineq2}.

%\red{The first inequality limits how large the upstream and downstream asymmetries can be simultaneously: an edge cannot be sampled arbitrarily strongly from the source side and be arbitrarily advantageous or harmful downstream unless the local excursion $n\to m\to n$ is correspondingly slow. The second inequality says that this effective local excursion time, when weighted by the steady traffic through the perturbed edge, is at most unity. Physically, one edge can renormalize the global first-passage clock by at most one effective local excursion.}

Eqs.~\eqref{mfptineq1}-\eqref{mfptineq2} together with the triangle inequality imply that the relative log-response is maximized at the perturbed point on the helpful side, and universally bounded by $1$ on the harmful side \footnote{We simply need to combine the exact expression for log-response,  $\frac{\partial\ln\tau_{kl}}{\partial \ln W_{mn}}=-\phi_{mn} (\tau_{kn}-\tau_{km})\left(1+\frac{\tau_{nk}-\tau_{nl}}{\tau_{kl}}\right)$, with these inequalities to obtain the desired relative log-response bound.}:
\begin{subequations}
\begin{align}
&\frac{\partial\ln\tau_{kl}}{\partial \ln W_{mn}}
\geq \frac{\partial\ln\tau_{mn}}{\partial \ln W_{mn}}
=-\phi_{mn}(\tau_{mn}+\tau_{nm})\geq -1,
\label{elasticitylower}\\
&\frac{\partial\ln\tau_{kl}}{\partial \ln W_{mn}}\leq \phi_{mn}\left(\tau_{nm}-\frac{(\tau_{kn}-\tau_{km})
       (\tau_{nk}-\tau_{nl})}{\tau_{kl}}\right) \leq 1.
\label{elasticityupper}
\end{align}
\end{subequations}
The inequality $\phi_{mn}(\tau_{mn}+\tau_{nm})\leq 1$ in the first line was proved in \cite{bao2024nonequilibrium}, and is
saturated when the edge $e_{mn}$ is a dynamical bottleneck. Notably, the bounds obtained here strengthen the unit response bounds derived in our companion paper~\cite{liang2026universal}. Eq.~\eqref{elasticitylower} applies when increasing $W_{mn}$ decreases the MFPT; it shows that the most negative relative response is realized locally at the perturbed point itself. Eq.~\eqref{elasticityupper} applies when increasing $W_{mn}$ increases the MFPT. Integrating these bounds along $W_{mn}\to cW_{mn}$ yields a coarser finite envelope $\min(c,c^{-1})\leq\tau_{kl}(cW_{mn})/\tau_{kl}(W_{mn})\leq \max(c,c^{-1})$. %while Eq.~\eqref{fullcurve} is sharper because it resolves the edge-specific nonlinear screening.

The same localization structure also applies to the bare MFPT response. By the triangle inequality,
\begin{subequations}
\begin{align}
\left|\frac{\partial\tau_{kl}}{\partial W_{mn}}\right|
&\leq \left|\frac{\partial\tau_{kn}}{\partial W_{mn}}\right|,
\label{local1}\\
\frac{\partial\tau_{kl}}{\partial W_{kn}}
&\leq \frac{\partial\tau_{kl}}{\partial W_{mn}},
\label{local2}
\end{align}
\end{subequations}
Equation~\eqref{local1} says that, for a fixed perturbed edge, the largest response magnitude is achieved when the initial state is the perturbed source side. Equation~\eqref{local2} says that, for a fixed MFPT, the strongest acceleration is obtained by perturbing the final step into the target. Together they show that the most negative MFPT responses (maximum acceleration) are local, namely,
\begin{equation}
    \min_{k,l,m,n}\frac{\partial\tau_{kl}}{\partial W_{mn}}
=\min_{m,n}\frac{\partial\tau_{mn}}{\partial W_{mn}}
=-\max_{m,n}\!\left\{\pi_n\tau_{mn}(\tau_{mn}+\tau_{nm})\right\}.
\label{local3_main}
\end{equation}

\textit{Higher-order response of MFPTs and steady-state probabilities---}The main result also generates analytic higher-order derivatives of MFPTs and steady-state probabilities: repeated differentiation of the linear response identities produces a
systematic nonlinear response hierarchy, expressed solely in terms of unperturbed MFPTs and steady-state probabilities. Such closed-form expressions were previously inaccessible.

As an illustration, we provide the second-order responses:
\begin{subequations}
    \begin{align}
  &\partial_X^2\pi_k=(2\alpha_X-1)\,\partial_X\pi_k,\\
  &\partial_{W_{mn}}^2\tau_{kl}=-2\pi_n\Sigma_{k|mn}\partial_{W_{mn}}\tau_{kl}.
\end{align}
\end{subequations}
Interestingly, the first derivative and curvature determine the nonlinear screening factors, $\pi_n\Sigma_{k|mn}$ for $\tau_{kl}$ and $\alpha_X$ for $\pi_n$, and hence reconstruct the full response curve for both the MFPT [combined with Eq.~\eqref{fullcurve}] and general steady-state or current observables under single-parameter perturbations. Here, $\alpha_X := \phi_{mn}\langle t_{nm}\rangle,\ \phi_{mn}\langle t_{nm}\rangle+\phi_{nm}\langle t_{mn}\rangle,\ \text{or}\ \pi_m  $ for $X=A_{mn},\ B_{mn},\ E_m$ respectively. For any state or current observable $\mathcal{O}$ in the steady-state, it has been shown that \cite{bao2024nonequilibrium} $\langle \mathcal{O}\rangle'=(1-e^{-\Delta X})\partial_{X}\langle \mathcal{O}\rangle/[1+(e^{-\Delta X}-1)\alpha_X]+\langle \mathcal{O}\rangle$, so determining the factor $\alpha_X$ is sufficient to determine the full response curve of $\langle \mathcal{O}\rangle$. %General mixed second derivatives are summarized in the End Matter.

\begin{figure}[!htb]
    \centering
    \includegraphics[width=0.98\columnwidth]{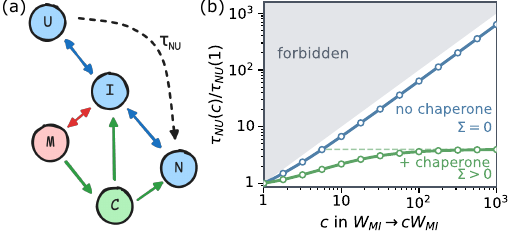}
    \caption{(a) Trap-bottleneck geometry and {the chaperone cycle acting downstream of} the same perturbed edge $I\to M$ (rose): {capture ($M\to C$), productive release ($C\to N$), and futile release ($C\to I$), all in green}. (b) Finite response of $\tau_{NU}$ to scaling the trap-entry rate $W_{MI}\to cW_{MI}$, on doubly logarithmic axes. The blue curve is the chaperone-free network, the green curve is the network with the chaperone, and the shaded region is forbidden by the universal harmful-side envelope $\tau(c)\leq c\,\tau(1)$. Solid lines are the exact prediction from Eq.~\eqref{nonlinearlinear}; markers are direct numerical computation. The trap is unscreened without the chaperone ($\Sigma_{N|MI}=0$) and screened with it ($\Sigma_{N|MI}>0$), so the response saturates at the plateau (dashed).}
    \label{fig:bio_response}
\end{figure}

% \begin{figure*}[t]
%     \centering
%     \includegraphics[width=0.98\textwidth]{fig1.png}
%     \caption{{(a) Trap-bottleneck geometry and two downstream modifications of the same perturbed edge $I\to M$: the rescue bypass (1, purple) and the productive shortcut (2, light blue). (b) Finite response of $\tau_{NU}$ to scaling the trap-entry rate $W_{MI}\to cW_{MI}$. The dark curve is the trap bottleneck, the dotted line is the universal harmful-side envelope $c$, the purple curve is the rescue bypass, and the light-blue curve is the productive shortcut. Solid lines are the exact prediction from Eq.~\eqref{nonlinearlinear}; markers are direct numerical computation. The trap bottleneck is harmful and unscreened ($\Sigma_{N|MI}=0$), the rescue bypass is harmful but screened ($\Sigma_{N|MI}>0$), and the productive shortcut makes the same perturbation helpful.}}
%     \label{fig:bio_response}
% \end{figure*}

\textit{Exact response for general multi-rate perturbations---}The linear response to a general multi-rate perturbation is directly obtained via the chain rule: $\frac{\partial \langle t_{kl}\rangle}{\partial \theta}=\sum_{m,n}\frac{\partial\langle t_{kl}\rangle}{\partial W_{mn}}\frac{\partial W_{mn}}{\partial \theta}.$ The multi-rate nonlinear response is more nontrivial. Using the exact finite single-rate update in End Matter together with Lemma~\ref{lemma1}, one can recursively obtain the MFPT after perturbations on more than one transition rate. This recursion yields exact analytic expressions for MFPTs and steady-state probabilities after perturbing any finite set of rates. The recursion is also computationally advantageous for sparse perturbations, and the corresponding update procedure and complexity estimates are summarized in End Matter.

\textit{Application I: Effect of bottleneck on response in a biologically-relevant folding network---}We illustrate the nonlinear response theory with a minimal folding network, see Fig. \ref{fig:bio_response}(a). The native folded state $N$ is the target, $U$ is the unfolded or uncommitted conformational ensemble from which folding starts, and the MFPT of interest, $\tau_{NU}$, naturally measures the mean folding time to the native state. $I$ is a productive intermediate, and the perturbed edge is the trap-entry transition $I\to M$ from the intermediate to an off-pathway misfolded state $M$, whose rate sets the misfolding propensity. The state $C$ is the chaperone--substrate complex: the chaperone captures the misfolded substrate ($M\to C$) and releases it productively ($C\to N$) or futilely ($C\to I$), the two limbs of one ATP-driven cycle \cite{delosrios2014hsp70,goloubinoff2018chaperones}. Such coarse-grained kinetic schemes are standard in protein folding and misfolding, where off-pathway traps and chaperone-assisted rescue routes shape folding times and proteostasis \cite{dobson2003protein,hartl2011molecular}. They also parallel the logic of perturbation analysis in biological Markov models, where one asks which kinetic step most strongly controls a completion time of interest \cite{caswell2011perturbation}.

Fig.~\ref{fig:bio_response}(b) isolates downstream nonlinear screening. Without the chaperone every path from $M$ to $N$ revisits $I$, so $\Sigma_{N|MI}=0$ and the response is affine in $c$ and unbounded; the chaperone opens $M\to C\to N$, which bypasses $I$, so $\Sigma_{N|MI}>0$ and the same perturbation saturates at a finite plateau. Additional numerical results separately illustrating upstream accessibility and downstream gain/loss are in \cite{supplemental_material}.

\textit{Application II: Dynamical shortcuts enhance target-resolved search efficiency in complex networks---}A natural %target-resolved 
measure of search efficiency in an $N$-state complex network is the global MFPT (GMFPT) to a target state $k$, defined as the average MFPT to $k$ over all possible initial states \cite{Benichou_globalmfpt}, i.e., $\overline{T}_k:=\sum_{l\neq k}\tau_{kl}/(N-1)$. Our response identity suggests a natural definition of a shortcut to the target $k$: an added or strengthened edge $e_{mn}$ from $n$ to $m$ is a dynamical shortcut if it moves trajectories to a state that is closer to $k$ in MFPT, namely $\tau_{km}<\tau_{kn}$. Adding such an edge from $n$ to $m$, the change in $\overline{T}_k$ is governed by
\begin{equation}
    \partial_{W_{mn}}\overline{T}_k=-S_{k,n}\pi_n (\tau_{kn}-\tau_{km}),
\end{equation}
where $S_{k,n}:=\sum_{l\neq k}(\tau_{kl}+\tau_{nk}-\tau_{nl})/(N-1)\geq 0$. Therefore, if $e_{mn}$ is a dynamical shortcut to $k$, then increasing its weight always decreases $\overline{T}_k$. In this target-resolved sense, a dynamical shortcut always enhances the search efficiency.

This clarifies the recently reported Braess-type paradox in Ref.~\cite{24prl_Braessparadox}. There, the search time used to quantify efficiency is not the target-resolved $\overline{T}_k$, but the further target-averaged quantity $\overline{T}:=\sum_k\overline{T}_k/N$. Our formula shows why $\overline{T}$ can indeed increase, as reported there: the same added edge can be a dynamical shortcut for some targets but not for others, because the sign of $\tau_{kn}-\tau_{km}$ is target dependent. Thus, the reported paradox mainly reflects the additional target averaging used to define the search-efficiency measure and is absent for the target-resolved GMFPT.

\textit{Discussion---}Using a novel correspondence between MFPT and steady-state probability response, we establish exact results on nonlinear responses for MFPTs and steady-state distributions to arbitrarily strong perturbations. This correspondence is the main conceptual step: it converts an
intrinsically transient response into a steady-state response problem in an auxiliary system, thereby
yielding exact MFPT response relations. Built on this key tool, Eqs.~\eqref{mfptlinear} and \eqref{nonlinearlinear} show that a global timing observable can be controlled by a small set of unperturbed MFPTs and steady-state probabilities, and that the response naturally separates into source-side accessibility, target-side gain/loss, and target-side nonlinear screening. This factorization provides a transparent physical picture of when an edge matters, why its finite response is suppressed, and how bottlenecks can be inferred from response data alone.

The framework also yields concrete practical consequences. A single nonlinear measurement, together with the unperturbed MFPT and its linear response, reconstructs the entire finite response curve. Recursive use of the exact single-rate update gives closed-form multi-rate perturbation formulas, sparse-update algorithms, and analytic higher-order derivatives. The folding example shows how upstream access, downstream gain/loss, and nonlinear screening can be separated in a biologically meaningful kinetic model, while the complex-network application shows that dynamical
shortcuts always enhance target-resolved search efficiency and explains how further target averaging
can obscure the shortcut notion and allow the previously reported paradox. %and the GMFPT application clarifies why target-resolved search efficiency avoids a Braess-type paradox.

Looking ahead, it will be natural to extend this response theory to higher-order moments and full distributions of first-passage times, potentially by combining it with matrix-calculus perturbation methods for absorbing chains \cite{caswell2011perturbation,caswell2019sensitivity}. It will also be important to test the bottleneck-screening decomposition in more realistic biological networks, including conformational switching, gene regulation, and disease-progression models.

\textit{Note added---}While finalizing this work, we became aware of the recent
work \cite{redner_24mfpt}. Although its title is similar in spirit, its focus is
complementary to ours: it studies the linear response of MFPTs to rare stochastic activations along trajectories, rather than deterministic perturbations of system parameters. 

%In contrast, we develop an exact response theory for MFPTs under finite perturbations of transition rates/system parameters

\begin{acknowledgments}
R.~B. was supported by JSPS KAKENHI Grant No. 25KJ0766. R. B. is grateful to Sosuke Ito and Naruo Ohga for discussions. S.~L. acknowledges financial support from the Max Planck Society. S.~L. thanks Pedro E. Harunari for an insightful discussion that inspired the auxiliary fast-reset approach.
\end{acknowledgments}

\bibliography{ref}

\appendix
\begin{center}{\large\bfseries{End Matter}}\end{center}

\begin{table*}[t!]
\centering
\caption{Summary of the procedure to obtain the closed-form expressions of new steady-state distributions and MFPTs under general multi-rate perturbations.}
\label{tab:procedure}

\renewcommand{\arraystretch}{1.35}
\setlength{\tabcolsep}{7pt}

\begin{tabularx}{\textwidth}{
|>{\centering\arraybackslash}m{0.06\textwidth}
|>{\centering\arraybackslash}X
|>{\centering\arraybackslash}m{0.55\textwidth}|
}
\hline
\textbf{Step} & \textbf{Description} & \textbf{Key expression} \\
\hline

1 &
Apply a perturbation $\Delta W_{mn}$ to a single transition rate $W_{mn}$. &
-- \\
\hline

2 &
Update the steady-state distribution using Lemma~1. &
\(\displaystyle
\pi^{\prime}_{k}
=
\pi_{k}
\left[
1+
\frac{(\tau_{kn}-\tau_{km})\Delta W_{mn}\pi_n}
{1+\Delta W_{mn}\pi_n\tau_{nm}}
\right]
\) \\
\hline

3 &
Compute the MFPT under the single perturbation using the updated steady-state distribution. &
\(\displaystyle
\frac{\tau_{kl}^{\prime}}{\tau_{kl}}
=
\frac{1}
{1+(\tau_{kn}-\tau_{km})\Delta W_{mn}\pi_n^{*\prime}}
\) \\
\hline

4 &
For multiple perturbations, e.g., on $W_{mn}$ and $W_{pq}$, express the MFPT recursively. &
\(\displaystyle
\frac{\tau_{kl}^{\prime\prime}}{\tau_{kl}}
=
\frac{\tau_{kl}^{\prime}/\tau_{kl}}
{1+(\tau_{kq}^{\prime}-\tau_{kp}^{\prime})\Delta W_{pq}\pi_q^{*\prime\prime}}
\) \\
\hline

5 &
Update the steady-state distribution for multiple perturbations using Lemma~1 and the finite MFPT expression. &
\(\displaystyle
\pi^{\prime\prime}_{k}
=
\pi_k^{\prime}
\left[
1+
\left(\tau_{kq}^{\prime}-\tau_{kp}^{\prime}\right)
\frac{\Delta W_{pq}\pi_q^{\prime}}
{1+\Delta W_{pq}\pi_q^{\prime}\tau_{qp}^{\prime}}
\right]
\) \\
\hline

6 &
Repeat the above steps recursively to obtain the exact expressions for both the steady-state distribution and the MFPTs under an arbitrary perturbation
$W\to W^{\prime}=W+\Delta W$. &
-- \\
\hline
\end{tabularx}
\end{table*}

\begin{table*}[t]
\centering
\footnotesize
\caption{Computational complexity comparison, assuming the benchmark MFPT matrix and stationary distribution are already known. The response-update column refers to a direct recursive implementation of Eq.~\eqref{finiteupdate} and Lemma~\eqref{lemma1}.}
\begin{ruledtabular}
\begin{tabular}{lccc}
Task & Direct recomputation & Response update & Advantage regime \\
\hline
All MFPTs and full $\pi'$ together & $O(N^4)$ & $O(KN^2)$ & $K\ll N^2$ \\
All MFPTs only & $O(N^4)$ & $O(KN^2)$ & $K\ll N^2$ \\
Full stationary vector $\pi'$ only & $O(N^3)$ & $O(K^2N)$ & $K\ll N$ \\
MFPTs to one fixed target & $O(N^3)$ & $O(K^2N)$ & $K\ll N$ \\
Full $\pi'$ after all required MFPT rows/columns are available & $O(N^3)$ & $O(KN)$ extra & always \\
\end{tabular}
\end{ruledtabular}
\label{tab:computational_advantage}
\end{table*}

\textit{Sketch of proof of the correspondence and the MFPT response identities---}To prove the exact single-rate response identities, we use a novel expression of the MFPT (see \cite{supplemental_material} for details):
\begin{equation}\label{newMFPT}
    \tau_{kl}=\frac{1}{\lim_{W_{lk}^{*}\to\infty} W_{lk}^{*}\pi_k^{*}},
\end{equation}
where $W_{lk}^{*}$ is the transition rate in a newly added unidirectional edge from state $k$ to state $l$. 

Notably, Eq.~\eqref{newMFPT} modifies Hill's original construction~\cite{hill2005free} (see also our companion paper~\cite{liang2026universal}), which redirects every transition into the target back to the initial state. The proposed expression is more transparent and is proved rigorously, whereas Hill's original construction~\cite{hill2005free} relied on intuitive arguments and provided no rigorous proof. More importantly, that construction can not yield the main result of this work, Eq.~\eqref{main_result}, because it simultaneously changes multiple edges and therefore does not permit a direct application of Lemma~1. The companion paper~\cite{liang2026universal} follows Hill's construction and accordingly does not obtain the exact response expressions expressed solely in terms of unperturbed quantities that are derived here. %The current expression is more transparent and is proven rigorously, while the original approach in \cite{hill2005free} was based on arguments. More importantly, we emphasize that the main result of this work, namely, Eq. \eqref{main_result} cannot be obtained from the original Hill's approach because it changes multiple edges so that Lemma 1 cannot be applied.

Therefore,
\begin{equation}\label{endmatter_keyidea}
    \frac{\tau_{kl}}{\tau_{kl}^{\prime}}=\lim_{W_{lk}^{*}\to\infty}\frac{\pi_k^{\prime*}}{\pi_k^{*}}=\lim_{W_{lk}^{*}\to\infty}\frac{\pi_k^{*\prime}}{\pi_k^{*}},
\end{equation}
where the first equality is from Eq.~\eqref{newMFPT}. The second equality follows from $\pi_k^{\prime *}=\pi_k^{*\prime}$, which holds because perturbing $W_{mn}$ and $W_{lk}$ in different orders results in the same perturbed transition-rate matrix. The key insight here is that the response of MFPT is equivalent to the response of a steady-state probability.

An application of the finite steady-state response identity $\pi_k^{\prime}-\pi_k=(\tau_{kn}-\tau_{km})\Delta W_{mn}\pi_n^{\prime}\pi_k$  yields (here we use the fact that $\tau^{*}_{kn}=\tau_{kn},\ \tau^{*}_{km}=\tau_{km}$, see Sec. II of \cite{supplemental_material})
\begin{equation}
    \frac{\pi_k^{*\prime}}{\pi_k^{*}}=(\tau_{kn}-\tau_{km})\Delta W_{mn}\pi_n^{*\prime}+1,
\end{equation}
which together with Eq.~\eqref{endmatter_keyidea} gives the exact formula of the perturbed MFPT: 
\begin{widetext}
    
\begin{align}\label{finiteupdate}
    \tau_{kl}^{\prime}
    &=\frac{\tau_{kl}}{(\tau_{kn}-\tau_{km})\Delta W_{mn}\pi_n^{*\prime}+1},\quad
    \pi^{*\prime}_n
    =\frac{\pi_n(\tau_{kl}+\tau_{nk}-\tau_{nl})}
        {\tau_{kl}(1+\tau_{nm}\pi_n\Delta W_{mn})
        -\pi_n\Delta W_{mn}(\tau_{kn}-\tau_{km})(\tau_{nk}-\tau_{nl})}.
\end{align}
\end{widetext}
Eq.~\eqref{finiteupdate} is the closed-form expression for the updated MFPT under an arbitrary single-rate perturbation. Here, $\pi_n^{*\prime}$ is solved by applying the multi-rate response identity derived in \cite{bao2024nonequilibrium,khodabandehlou2024Affine}
\begin{equation}
    \pi_k^{\prime}-\pi_k=\sum_{m<n}(\tau_{kn} - \tau_{km})(\Delta W_{mn}\pi_n^{\prime}-\Delta W_{nm}\pi_m^{\prime})\pi_k.
\end{equation}

Further algebraic manipulation of Eq.~\eqref{finiteupdate} yields the nonlinear response identity Eq.~\eqref{nonlinear_identity}, with details in \cite{supplemental_material} Sec. II. Taking the linear-response limit ($\Delta W_{mn}\to 0$) leads to the identities Eqs.~\eqref{mfptlinear} and ~\eqref{nonlinearlinear}. Eq.~\eqref{mfptlinear} can also be directly obtained from the linear-response correspondence $\partial_{W_{mn}}\ln \tau_{kl}=-\lim_{W_{lk}^{*}\to\infty}\partial_{W_{mn}}\ln\pi_k^*$. 

\textit{Procedure for recursive multi-rate updates---}The exact single-rate update in Eq.~\eqref{finiteupdate}, together with Lemma~\ref{lemma1} can be applied recursively. Starting from the original network, one first updates the steady-state distribution under one perturbed rate, then uses those updated MFPTs and steady-state probabilities as the input for the next perturbed rate. Repeating this procedure yields exact closed-form expressions for both MFPTs and steady-state probabilities under any finite set of rate perturbations. Table~\ref{tab:procedure} summarizes the recursion.

\textit{Advantage in computational complexity---}From a computational perspective, our framework offers substantial numerical advantages. Having derived closed-form expressions for both perturbed MFPTs and stationary distributions, we can leverage known information from any benchmark system to efficiently compute MFPTs and stationary distributions for any other systems with the same number of states as the benchmark. The relevant cost is controlled by the number $K$ of perturbed transition rates and by the dependency closure generated by the recursive update.

\end{document}

\begin{CJK*}{UTF8}{gbsn}
\title{Supplementary Material for ``Exact First-Passage Time Response Theory from Steady-State Response''
}

\author{Ruicheng Bao}
\email{Corresponding author: ruicheng@g.ecc.u-tokyo.ac.jp}
\affiliation{Department of Physics, Graduate School of Science, 
The University of Tokyo, Hongo, Bunkyo-ku, Tokyo 113-0033, Japan}

\author{Shiling Liang (梁师翎)}
\email{Corresponding author: shiling@pks.mpg.de}
\affiliation{Center for Systems Biology Dresden, 01307 Dresden, Germany}
\affiliation{Max Planck Institute for the Physics of Complex Systems, 01187 Dresden, Germany}
\affiliation{Max Planck Institute of Molecular Cell Biology and Genetics, 01307 Dresden, Germany}
\maketitle
\end{CJK*}

\tableofcontents

\noindent {Throughout this Supplemental Material, we define $\tau_{kl}$ as the MFPT from state $l$ to state $k$.}

\section{Review of important properties of MFPT}

\paragraph{Kemeny's constant}
For any two states $m$ and $n$:
\begin{equation}
    \sum_k \pi_k\tau_{km} =  \sum_k \pi_k\tau_{kn},
\end{equation}
which is a constant in the sense  that it is irrelevant to the initial state.

\paragraph{Recursive relation of MFPTs}
For any states $m$ and $n$ ($m\neq n$), it holds that:
\begin{equation}\label{eq:mfpt_orig}
    \tau_{mn} = -\frac{1}{W_{nn}} - \sum_{l\neq n}\frac{W_{ln}}{W_{nn}} \tau_{ml},
\end{equation}
where $W_{ln}$ represents the transition rate from state $n$ to state $l$. The physical interpretation of this recursive relation is clear: First, the term $-1/W_{nn}$ represents the average time spent in state $n$ before any transition occurs---that is, the mean escape time from state $n$. Second, when leaving state $n$, the process jumps to another state $l$ with a probability $W_{ln}/(-W_{nn})$ and then requires an additional mean time $\tau_{ml}$ to reach state $m$ (note that when $m=l$, the additional time $\tau_{mm}=0$). Thus, the total MFPT $\tau_{mn}$ is simply the sum of the waiting time in $n$ plus the weighted contributions of the times required to go from the subsequent states to $m$. If $m=n$, the $\tau_{mm} $ should be replaced with the mean first return time $T_{mm}$ of state $m$, i.e.,
\begin{equation}\label{return}
    T_{mm} = -\frac{1}{W_{mm}} - \sum_{l\neq n}\frac{W_{lm}}{W_{mm}} \tau_{ml}.
\end{equation}
Here, the mean first return time $T_{mm}$ is given by
\begin{equation}
    T_{mm}=\frac{1}{-W_{mm}\pi_m}.
\end{equation}

\paragraph{Triangle inequality of MFPTs}
For any three states $k,m,n$, there is a triangle inequality 
\begin{equation} \label{triangle}
    \tau_{km} +\tau_{mn} \geq \tau_{kn},
\end{equation}
with the equality condition that all first-passage paths from state $n$ to state $k$ go through the state $m$.

\section{Proof of the main results}
Here, we would like to the theorem below, which can generate all the main results of the main text, namely, the linear response and nonlinear response of an arbitrary MFPT:

\textit{Theorem 1---}
The response of the MFPT from an arbitrary state $l$ to another state $k$ to the perturbation $W_{mn}\to W_{mn}+\Delta W_{mn}$ ($\Delta W_{mn}\geq -W_{mn}$) can be exactly expressed using the MFPTs and the steady state distribution of the unperturbed dynamics:
\begin{align}
     &\langle t_{kl}\rangle^{\prime}=\frac{\langle t_{kl}\rangle}{(\langle t_{kn}\rangle - \langle t_{km}\rangle)\Delta W_{mn}\pi_n^{*\prime}+1}, \label{MFPTfinite}\\ 
     &\text{with}\quad\pi^{* \prime}_n = \frac{\pi_n\Bigl(\langle t_{kl}\rangle +\langle t_{nk}\rangle-\langle t_{nl}\rangle\Bigr)}
        {\langle t_{kl}\rangle\Bigl(1+\langle t_{nm}\rangle\,\pi_n\,\Delta W_{mn}\Bigr)
        -\pi_n\,\Delta W_{mn}\Bigl(\langle t_{kn}\rangle-\langle t_{km}\rangle\Bigr)
        \Bigl(\langle t_{nk}\rangle-\langle t_{nl}\rangle\Bigr)}.\nonumber
\end{align}
To prove this theorem, we first propose some prerequisites in what follows.

\textit{Prerequisite 1: A novel expression of the MFPT}

To prove theorem 1, we shall use a novel expression of the MFPT developed in the Appendix A, which is based on a modified Hill's approach:
\begin{equation}\label{newMFPT}
    \tau_{kl} =\frac{1}{\lim_{W_{lk}^{*}\to\infty} W_{lk}^{*}\pi_k^{*}}, 
\end{equation}
where $W_{lk}^{*}$ is the transition rate in a newly added unidirectional edge from state $k$ to state $l$ (the $\lim_{W_{lk}^{*}\to\infty}$ is omitted in what follows for brevity). This is proven in what follows.

Physically, if the continuous-time Markov system is instantaneously reset to its the initial state $i$ when it reach the target state $j$, then its average number of transition from state $j$ to state $i$ within a duration $\tau$, $N_{i\to j}^{\tau}$ , would be the average number of first-passage event from state $i$ to state $j$ in that duration. Then, the MFPT is given by $\tau_{ji} = \frac{\tau}{N_{i\to j}^{\tau}}$, i.e., the average number of transition from state $j$ to $i$ per time unit in this setting. This is the physical idea behind Hill's original approach. Here, we realize this idea using a modified approach: we realize the instantaneous reset by adding a one-way transition from state $j$ to state $i$ with the new transition rate $W_{ij}^{*}\to\infty$. In Hill's original approach, the reset is realized by redirecting all transitions into $j$ to state $i$, and eliminate the target state $j$. In our approach, it is clear that the average number of transitions from $j$ to $i$ per time unit is $\frac{1}{\phi_{ij}^{*}}=\frac{1}{W_{ij}^{*}\pi_{j}^{*}}$,
where $\pi_{j}^{*}\rightarrow0$ is the steady state probability of $j$ in the modified network with a new edge $W_{ij}^{*}$. Using Lemma 1, we can mathematically prove that the relation $\tau_{ji}=1/\phi_{ij}^{*}$ indeed holds:
\begin{align}\label{limitpi}
    \phi_{ij}^{*}&=\lim_{W_{ij}^{*}\to\infty}(W_{ij}^{0}+W_{ij}^{*})\pi_j\left(1- \frac{\tau_{ji} W_{ij}^{*}\pi_j}{1+\tau_{ji} W_{ij}^{*}\pi_j} \right)\\
    &=\lim_{W_{ij}^{*}\to\infty}\frac{(W_{ij}^{0}+W_{ij}^{*})\pi_j}{1+\tau_{ji} W_{ij}^{*}\pi_j}=\frac{1}{\tau_{ji}}.
\end{align}

A byproduct: an arbitrary steady-state current $j_{kl}:=\phi_{kl}-\phi_{lk}$ is the monotonically increasing (decreasing) function of $W_{kl}$ ($W_{lk}$). Thus, the upper bound and lower bound for the current can be identified using the novel expression of MFPT as 
\begin{equation}
    -\frac{1}{\tau_{kl}}\leq j_{kl}\leq \frac{1}{\tau_{lk}},
\end{equation}
because it is easy to check $\lim_{W_{lk}\to\infty} j_{kl}=-1/\tau_{kl}$ and $\lim_{W_{kl}\to\infty} j_{kl}=1/\tau_{lk}$.

\textit{Prerequisite 2: Nonlinear response of steady-state probability to strong perturbation}

Consider the general perturbation on all transition rates, we have proven in \cite{bao2024nonequilibrium} that
\begin{equation}
    \pi_k^{\prime}-\pi_k=\sum_{m<n}(\tau_{kn} - \tau_{km})(\Delta W_{mn}\pi_n^{\prime}-\Delta W_{nm}\pi_m^{\prime})\pi_k.\label{Fmfpt}
\end{equation}

We also have a duality equation:
\begin{equation}
    \pi_k^{\prime}-\pi_k=\sum_{m<n}(\tau_{kn}^{\prime} - \tau_{km}^{\prime})(\Delta W_{mn}\pi_n-\Delta W_{nm}\pi_m)\pi_k^{\prime}.\label{Fmfpt2}
\end{equation}

With the expression of MFPT, we have 
\begin{equation}
    \frac{\tau_{kl}}{\tau_{kl}^{\prime}}=\lim_{\Delta W_{lk}\to\infty}\frac{\pi_k^{\prime*}}{\pi_k^{*}}=\lim_{\Delta W_{lk}\to\infty}\frac{\pi_k^{*\prime}}{\pi_k^{*}}.
\end{equation}
The key insight here is the equality $\pi_k^{\prime *}=\pi_k^{*\prime}$, which holds because the order of perturbations to $W_{mn}$ and $W_{lk}$ is irrelevant---the resulting transition rate matrix remains the same regardless of the perturbation sequence. The first equality is from Eq. \eqref{newMFPT}. We note that the first equality can also be directly obtained by applying Lemma 1 twice, first for the perturbation on $\Delta W_{lk}\to \infty$ for $\pi_k^{\prime}$, then for $\pi_k$, that is,
\begin{equation}
    \lim_{\Delta W_{lk}\to\infty}\frac{\pi_k^{\prime*}}{\pi_k^*}=\lim_{\Delta W_{lk}\to\infty}\frac{\frac{\pi_k^{\prime}}{1+\Delta W_{lk}\pi_k^{\prime}\tau_{kl}^{\prime}}}{\frac{\pi_k}{1+\Delta W_{lk}\pi_k\tau_{kl}}}=\lim_{\Delta W_{lk}\to\infty}\frac{\Delta W_{lk}\pi_k\pi_k^{\prime}\tau_{kl}}{\Delta W_{lk}\pi_k^{\prime}\pi_k\tau_{kl}^{\prime}}=\frac{\tau_{kl}}{\tau_{kl}^{\prime}}.
\end{equation}

We then apply Eq. \eqref{Fmfpt} to the new steady-probability $\pi_k^{*}$. If there is a finite perturbation $\Delta W_{mn}$ on $W_{mn}$, the response of $\pi_k^{*}$ is given by
\begin{equation}
    \frac{\pi_k^{*\prime}}{\pi_k^{*}}=(\tau_{kn}^*-\tau_{km}^*)\Delta W_{mn}\pi_n^{*\prime}+1.
\end{equation}
We observe that the new edge $W_{lk}$ is out of state $k$, so any first-passage event targeted to state $k$ will not be affected by the change in this edge. Consequently, and $\tau_{ki}^*=\tau_{ki}$ for any state $i$. 
Therefore,
\begin{equation*}
    \tau_{kn}^*-\tau_{km}^*=\tau_{kn}-\tau_{km}
\end{equation*} 
so that 
\begin{equation}
    \frac{\pi_k^{*\prime}}{\pi_k^{*}}=(\tau_{kn}-\tau_{km})\Delta W_{mn}\pi_n^{*\prime}+1.
\end{equation}

The last step to finish the proof of Theorem 1 is solving $\pi^{* \prime}_n$. This can be achieved by solving the system of two linear equations,
\[
\left\{
\begin{aligned}
\frac{\pi''_k - \pi_k}{\pi_k} &= [(\tau_{kn} - \tau_{km})\Delta W_{mn}\,\pi''_n 
- \tau_{kl}\,\Delta W_{lk}\,\pi''_k],\\[1mm]
\frac{\pi''_n - \pi_n}{\pi_n} &= [(\tau_{nk} - \tau_{nl})\Delta W_{lk}\,\pi''_k 
- \tau_{nm}\,\Delta W_{mn}\,\pi''_n].
\end{aligned}
\right.
\]
and taking the $\Delta W_{lk}\rightarrow \infty$ limit, i.e. $\pi^{* \prime}_n=\lim_{\Delta W_{lk} \to \infty}\pi_n^{\prime\prime}$. This system of two linear equations is a consequence of \eqref{Fmfpt}. The solution is 
\begin{equation}\label{exact_pi}
    \pi^{* \prime}_n = \frac{\pi_n\Bigl(\tau_{kl} +\tau_{nk}-\tau_{nl}\Bigr)}
        {\tau_{kl}\Bigl(1+\tau_{nm}\,\pi_n\,\Delta W_{mn}\Bigr)
        -\pi_n\,\Delta W_{mn}\Bigl(\tau_{kn}-\tau_{km}\Bigr)
        \Bigl(\tau_{nk}-\tau_{nl}\Bigr)}.
\end{equation}
We thus complete the proof. $\square$

\subsection{From the exact finite update to the nonlinear--linear identity}
We now use Theorem 1 to prove the nonlinear-linear response identity in the main text. Starting from the exact finite update,
{
\begin{equation}
    \tau_{kl}^{\prime}=\frac{\tau_{kl}}{1+(\tau_{kn}-\tau_{km})\Delta W_{mn}\pi_n^{* \prime}},
\end{equation}}
we substitute Eq.~\eqref{exact_pi} into the denominator and obtain
{
\begin{align}
&1+(\tau_{kn}-\tau_{km})\Delta W_{mn}\pi_n^{* \prime} \nonumber\\
&=1+\frac{(\tau_{kn}-\tau_{km})\Delta W_{mn}\pi_n(\tau_{kl}+\tau_{nk}-\tau_{nl})}{\tau_{kl}(1+\tau_{nm}\pi_n\Delta W_{mn})-\pi_n\Delta W_{mn}(\tau_{kn}-\tau_{km})(\tau_{nk}-\tau_{nl})} \nonumber\\
&=\frac{\tau_{kl}(1+\tau_{nm}\pi_n\Delta W_{mn})-\pi_n\Delta W_{mn}(\tau_{kn}-\tau_{km})(\tau_{nk}-\tau_{nl})}{\tau_{kl}(1+\tau_{nm}\pi_n\Delta W_{mn})-\pi_n\Delta W_{mn}(\tau_{kn}-\tau_{km})(\tau_{nk}-\tau_{nl})} \nonumber\\
&\qquad+\frac{(\tau_{kn}-\tau_{km})\Delta W_{mn}\pi_n(\tau_{kl}+\tau_{nk}-\tau_{nl})}{\tau_{kl}(1+\tau_{nm}\pi_n\Delta W_{mn})-\pi_n\Delta W_{mn}(\tau_{kn}-\tau_{km})(\tau_{nk}-\tau_{nl})} \nonumber\\
&=\frac{\tau_{kl}+\tau_{kl}\pi_n\Delta W_{mn}(\tau_{nm}+\tau_{kn}-\tau_{km})}{\tau_{kl}(1+\tau_{nm}\pi_n\Delta W_{mn})-\pi_n\Delta W_{mn}(\tau_{kn}-\tau_{km})(\tau_{nk}-\tau_{nl})}.
\end{align}}
Therefore,
{
\begin{equation}\label{update_MPFT}
    \tau_{kl}^{\prime}=\frac{\tau_{kl}(1+\tau_{nm}\pi_n\Delta W_{mn})-\pi_n\Delta W_{mn}(\tau_{kn}-\tau_{km})(\tau_{nk}-\tau_{nl})}{1+\pi_n\Delta W_{mn}(\tau_{nm}+\tau_{kn}-\tau_{km})}.
\end{equation}}
Subtracting $\tau_{kl}$ from both sides gives
{
\begin{align}
\tau_{kl}^{\prime}-\tau_{kl}
&=\frac{\tau_{kl}(1+\tau_{nm}\pi_n\Delta W_{mn})-\pi_n\Delta W_{mn}(\tau_{kn}-\tau_{km})(\tau_{nk}-\tau_{nl})}{1+\pi_n\Delta W_{mn}(\tau_{nm}+\tau_{kn}-\tau_{km})}-\tau_{kl} \nonumber\\
&=-\frac{\pi_n\Delta W_{mn}(\tau_{kn}-\tau_{km})(\tau_{kl}+\tau_{nk}-\tau_{nl})}{1+\pi_n\Delta W_{mn}(\tau_{nm}+\tau_{kn}-\tau_{km})}.
\end{align}}
Using the linear-response formula of the main text,
{
\begin{equation}
    \partial_{W_{mn}}\tau_{kl}=-\pi_n(\tau_{kn}-\tau_{km})(\tau_{kl}+\tau_{nk}-\tau_{nl}),
\end{equation}}
we finally obtain the exact nonlinear--linear identity
{
\begin{equation}
    \frac{\tau_{kl}^{\prime}-\tau_{kl}}{\Delta W_{mn}}=\frac{\partial_{W_{mn}}\tau_{kl}}{1+\pi_n\Delta W_{mn}(\tau_{nm}+\tau_{kn}-\tau_{km})}.
\end{equation}}

\subsection{Numerical verification of the exact finite single-rate update}
\begin{figure}[!htb]
    \centering
    \includegraphics[width=0.4\textwidth]{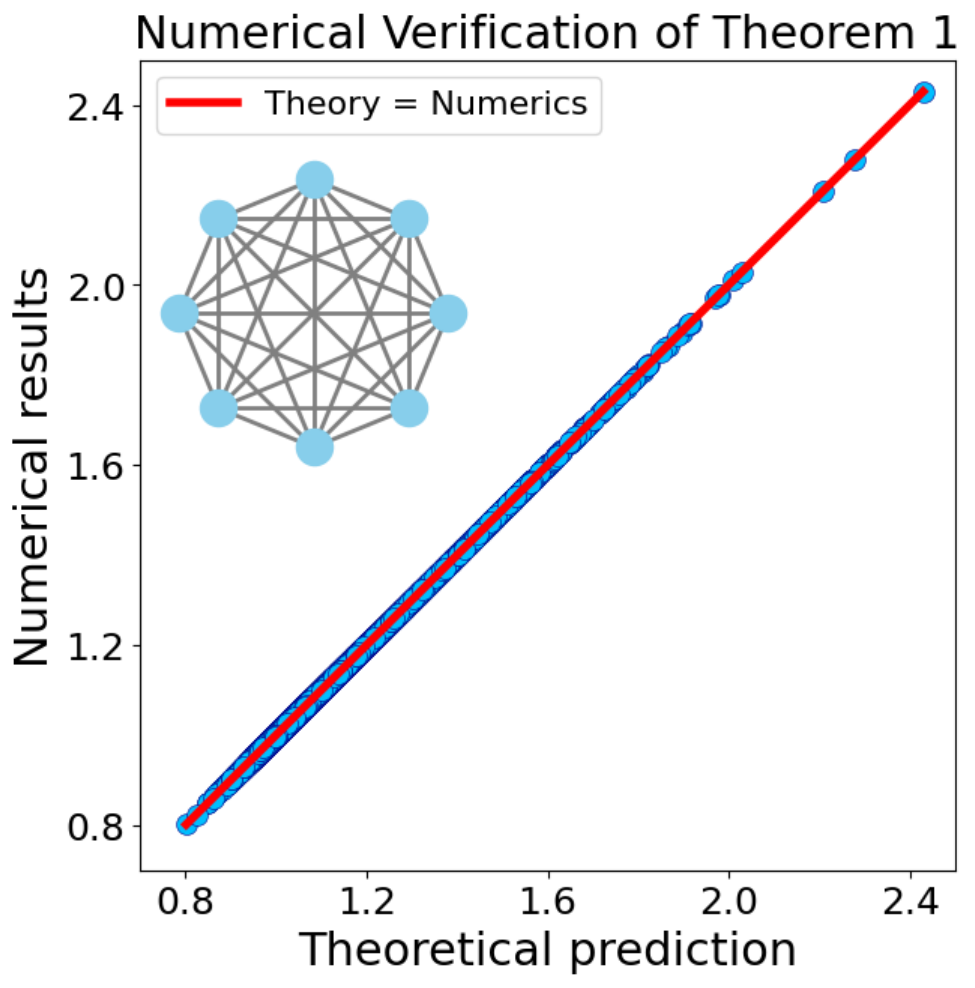}
    \caption{{Numerical verification of the exact finite single-rate update (Theorem~1). In each of the $10^5$ iterations, an 8-state fully connected Markov chain is randomly generated with transition rates drawn uniformly from $[0.01,10]$. One pair $(k,l)$ is chosen for the MFPT, another pair $(m,n)$ for the perturbation, and the perturbation strength is drawn uniformly from $[0.01,10]$. The numerical MFPT ratio agrees with the exact prediction.}}
    \label{fig:sm_theorem1}
\end{figure}

A direct numerical verification of Theorem 1 is shown in Fig.~\ref{fig:sm_theorem1}.

\textit{An alternative proof of the response to local perturbation}

We can obtain the response to local perturbation solely using Eqs. \eqref{Fmfpt}-\eqref{Fmfpt2}. Consider a perturbation on a single rate $W_{ij}$, Eqs. \eqref{Fmfpt}-\eqref{Fmfpt2} reduce to
\begin{align}
    &\pi_i^{\prime}-\pi_i=\tau_{ij} \Delta W_{ij}\pi_j^{\prime}\pi_i=\tau_{ij}^{\prime} \Delta W_{ij}\pi_j\pi_i^{\prime}\\
    &\pi_j^{\prime}-\pi_j=-\tau_{ji} \Delta W_{ij}\pi_j^{\prime}\pi_j=-\tau_{ji}^{\prime}\Delta W_{ij}\pi_j\pi_j^{\prime},
\end{align}
Defining $\Delta \pi_i=\pi_i^{\prime}-\pi_i$ for brevity, we get
\begin{align}
    &\tau_{ij}=\frac{\Delta \pi_i}{\Delta W_{ij}}\frac{1}{\pi_j^{\prime}\pi_i},\quad \tau_{ij}^{\prime}= \frac{\Delta \pi_i}{\Delta W_{ij}}\frac{1}{\pi_j\pi_i^{\prime}}\label{MFPT_response1}\\
    &\tau_{ji}=-\frac{\Delta \pi_j}{\Delta W_{ij}}\frac{1}{\pi_j^{\prime}\pi_j},\quad \tau_{ji}^{\prime}= -\frac{\Delta \pi_i}{\Delta W_{ij}}\frac{1}{\pi_j\pi_j^{\prime}}.\label{MFPT_response2}
\end{align}
Similarly, for a perturbation on $W_{ji}$, we have
\begin{align}
    &\pi_i^{\prime}-\pi_i=-\tau_{ij} \Delta W_{ji}\pi_i^{\prime}\pi_i=\tau_{ij}^{\prime}\Delta W_{ji}\pi_i\pi_i^{\prime},\\
    &\pi_j^{\prime}-\pi_j=\tau_{ji} \Delta W_{ji}\pi_i^{\prime}\pi_j=\tau_{ji}^{\prime} \Delta W_{ji}\pi_i\pi_j^{\prime}
\end{align}

The response of MFPT to finite perturbation on $W_{ij}$ is given by
\begin{equation}
\begin{aligned}
     \tau_{ij}^{\prime}-\tau_{ij} &=\frac{\Delta\pi_i}{\Delta W_{ij}} \left(\frac{1}{\pi_j \pi_i^{\prime}}-\frac{1}{\pi_j^{\prime}\pi_i}\right)\\
    &=\frac{\Delta\pi_i}{\Delta W_{ij}} \left(\frac{(\pi_j^{\prime}\pi_i-\pi_j\pi_i)- (\pi_j \pi_i^{\prime}-\pi_j\pi_i)}{\pi_i \pi_j \pi_i^{\prime}\pi_j^{\prime}}\right)\\
    &=\frac{\Delta\pi_i}{\Delta W_{ij}} \frac{\pi_i \Delta \pi_j-\pi_j\Delta \pi_i}{\pi_i \pi_j \pi_i^{\prime}\pi_j^{\prime}}.
\end{aligned}
\end{equation}
Using Eqs. \eqref{MFPT_response1}-\eqref{MFPT_response2}, we find
\begin{equation}
    \frac{\tau_{ij}^{\prime}-\tau_{ij}}{\Delta W_{ij}}=-\frac{\Delta\pi_i}{\Delta W_{ij}} \frac{\tau_{ij}+\tau_{ji}}{\pi_i^{\prime}}=-\tau_{ij}^{\prime}\pi_j(\tau_{ij}+\tau_{ji}).
\end{equation}
Similarly, when $W_{ji}$ is perturbed,
\begin{equation}
    \tau_{ij}^{\prime}-\tau_{ij}=\frac{\Delta\pi_i}{\Delta W_{ji}} \left( \frac{1}{\pi_i\pi_i^{\prime}}-\frac{1}{\pi_i^{\prime}\pi_i}\right)=0.
\end{equation}

\section{Sign of $\tau_{kn}-\tau_{km}$ determines the sign of $\partial_{W_{mn}}\tau_{kl}$}
Here we provide another interpretation regarding the sign of the MFPT response. The MFPT from $l$ to $k$ can be formally expressed as
\begin{equation}\label{formal}
    \tau_{kl}=(\tau_{S}+\tau_{nl})p_n+(1-p_n)\tau_{l\to k}^{\setminus n},
\end{equation}
where $p_n$ is the probability that an arbitrary first passage trajectory from $l$ to $k$ including state $n$ and $\tau_{l\to k}^{\setminus n}$ is the average time over first-passage events not going through $n$. Here, $\tau_{nl}$, $p_n$ and $\tau_{l\to k}^{\setminus n}$ are all independent of $W_{mn}$. Thus, the only contribution of the response of $\tau_{kl}$ to $W_{mn}$ is from $\tau_{S}=\tau_{kn}$. Intuitively, when $W_{mn}\to W_{mn}+\Delta W_{mn}$ with $\Delta W_{mn}>0$, the perturbed $\tau_{S}$, denoted as $\tau_{S}^{\prime}=\tau_{kn}^{\prime}$, will be closer to $\tau_{km}^{\prime}$. To gain further physical intuition, we consider the extreme case when $\Delta W_{mn}\to\infty$. In this case, $\tau_{S}^{\prime}=\tau_{km}^{\prime}=\tau_{kn}^{\prime}$, i.e., two MFPTs converge to the same value as $W_{mn}$ increases. Thus, if $\tau_{kn}>\tau_{km}$, the value of $\tau_{S}=\tau_{kn}$ will become smaller to get closer to $\tau_{km}$ when increasing $W_{mn}$. According to \eqref{formal}, this implies that $\tau_{kl}$ decreases, i.e., negatively responds to the change of $W_{mn}$. If $\tau_{kn}<\tau_{km}$, the value of $\tau_{kl}$ will increase as $W_{mn}$ increases, following the same logic. The above discussion explains why the sign of $\partial_{W_{mn}}\tau_{kl}$ is determined by the sign of $\tau_{kn}-\tau_{km}$.

\section{Any MFPT $\tau_{kl}$ is a monotonic function of an arbitrary transition rate $W_{mn}$}
Recall that the linear response of an arbitrary MFPT to a single-rate perturbation on $W_{mn}$ is given by 
\begin{equation}\label{linearrecall}
     \frac{\partial\tau_{kl}}{\partial W_{mn}}=-\pi_n (\tau_{kn}-\tau_{km})(\tau_{kl}+\tau_{nk}-\tau_{nl}).
\end{equation}
Therefore, to prove the monotonicity of $\tau_{kl}$ with respect to $W_{mn}$ is equivalent to prove that either $\tau_{kn}-\tau_{km}\geq 0$ or $\tau_{kn}-\tau_{km}\leq 0$ hold for any value of $W_{mn}$. That is, $ \frac{\partial\tau_{kl}}{\partial W_{mn}}\geq 0$ for any $W_{mn}$ ($\tau_{kl}$ increases monotonically as $W_{mn}$ increases) or vice versa. 

By choosing $l=m$ and $l=n$ in Eq. \eqref{linearrecall}, it is straightforward to show that
\begin{equation}
    \frac{\partial(\tau_{kn} -\tau_{km})}{\partial W_{mn}}=-\pi_n(\tau_{kn} -\tau_{km})(\tau_{kn}+\tau_{nm}-\tau_{km}).
\end{equation}
Since $\tau_{kn}+\tau_{nm}-\tau_{km}\geq 0$ due to the triangle inequality \eqref{triangle}, we have
\begin{equation}
    \frac{1}{\tau_{kn} -\tau_{km}}\frac{\partial(\tau_{kn} -\tau_{km})}{\partial W_{mn}}=-\pi_n(\tau_{kn}+\tau_{nm}-\tau_{km})\leq 0
\end{equation}
for any $W_{mn}$. 

Let $f(W_{mn}) = \tau_{kn} - \tau_{km}$. Typically, $f(W_{mn})$ is a continuous function of $W_{mn}$. We know that   
$\frac{f'(W_{mn})}{f(W_{mn})} \leq 0$ for all $W_{mn}$. Suppose, for contradiction, that $\frac{\partial(\tau_{kn} -\tau_{km})}{\partial W_{mn}}$ changes sign at some point. This means there exists a value $W_{mn}^*$ where either:  

(i) $f(W_{mn}^*) = 0$ with $f'(W_{mn}^*) \neq 0$, or  
(ii) $f'(W_{mn}^*) = 0$ with $f(W_{mn}^*) = 0$, and $f$ changes sign across $W_{mn}^*$.  

In case (i), $f(W_{mn})$ changes sign across $W_{mn}^*$ while $f'(W_{mn})$ maintains its sign. This creates an interval where $f(W_{mn})$ and $f'(W_{mn})$ have the same sign, violating $\frac{f'(W_{mn})}{f(W_{mn})} \leq 0$.  

In case (ii), if $f(W_{mn})$ changes sign across $W_{mn}^*$ (odd-order zero), then similarly, there must be an interval where $f(W_{mn})$ and $f'(W_{mn})$ have the same sign.  

Therefore, $f(W_{mn})$ must maintain a consistent sign throughout its domain, which implies that $\frac{\partial(\tau_{kn} -\tau_{km})}{\partial W_{mn}}$ must be either always non-positive or always non-negative ($\tau_{kn} -\tau_{km}$ should be either non-negative or non-positive). 

The above argument proves that $\frac{\partial\tau_{kl}}{\partial W_{mn}}$ is either always non-negative or always non-positive, i.e., $\tau_{kl}$ is monotonic with respect to $W_{mn}$.    

\section{Convexity and Concavity of MFPT}
From the relation
\begin{equation}
    \frac{\tau_{kl}^{\prime}-\tau_{kl}}{\Delta W_{mn}}=\frac{\frac{\partial \tau_{kl}}{\partial W_{mn}}}{1+\Delta W_{mn}\pi_n(\tau_{nm}+\tau_{kn} -\tau_{km})},
\end{equation}
we conclude that the inequality
\begin{equation}
     \left|\frac{\tau_{kl}^{\prime}-\tau_{kl}}{\Delta W_{mn}}\right|\leq \left|\frac{\partial \tau_{kl}}{\partial W_{mn}} \right|
\end{equation}
holds for any value of $W_{mn}$ (assuming $\Delta W_{mn}>0$ without loss of generality). Taylor expanding $\tau_{kl}^{\prime}-\tau_{kl}$ to second order, it is straightforward to show that $\partial_{W_{mn}}\tau_{kl}\geq 0$ implies $\partial^2_{W_{mn}^2}\tau_{kl}\leq 0 $ (concavity) and $\partial_{W_{mn}}\tau_{kl}\leq 0$ implies $\partial^2_{W_{mn}^2}\tau_{kl}\geq 0 $ (convexity)

\section{Proof of some results in the main text}
Here, we provide proofs of some corollaries in the main text using the Kemeny's constant and the recursion relation of MFPTs.

\subsection{Corollaries about the energy-perturbation and the energy-barrier perturbation}

\paragraph{Response to energy-perturbation}

We first recall that the transition rate is parametrized as 
\begin{equation}
    W_{nm}=e^{-B_{nm}+E_m+f_{nm}}.
\end{equation}
Then, the response to energy perturbation on $E_m$ can be calculated by Eq. \eqref{linearrecall} and the chain-rule as
\begin{align}
    \frac{\partial  \tau_{kl}}{\partial E_m}&=\sum_n \frac{\partial\tau_{kl}}{\partial W_{nm}}\frac{\partial W_{nm}}{\partial E_m}=\sum_n\frac{\partial\tau_{kl}}{\partial W_{nm}}W_{nm}\\
    &=-\sum_n W_{nm} \pi_m (\tau_{km}-\tau_{kn})(\tau_{kl}+\tau_{mk}-\tau_{ml})\\
    &=-\pi_m \left[\left(\sum_{n (\neq m)} W_{nm}\right)\tau_{km}-\left(\sum_{n (\neq m)} W_{nm}\tau_{kn}\right)\right](\tau_{kl}+\tau_{mk}-\tau_{ml})\\
    &=-\pi_m \left[\left| W_{mm}\right|\tau_{km}-\left( |W_{mm}|\tau_{km}-1\right)\right](\tau_{kl}+\tau_{mk}-\tau_{ml})\\
    &=-\pi_m(\tau_{kl}+\tau_{mk}-\tau_{ml}),
\end{align}
which proves the corollary 
\begin{equation*}  
\frac{\partial  \tau_{kl}}{\partial E_m} = -\pi_m(\tau_{kl}+\tau_{mk}-\tau_{ml})\leq 0
\end{equation*}
in the main text. 

The result can also be proven by noticing that
\begin{equation}
    \frac{\partial  \ln\tau_{kl}}{\partial E_m}=-\frac{\partial \ln \pi_k^{*}}{\partial E_m}=-\pi_m^{*},
\end{equation}
where $\pi_k^*=\lim_{W_{lk}\to\infty}\pi_k$ and $$\pi_m^*=\lim_{W_{lk}\to\infty}\pi_m=\pi_m\left(1+\frac{\tau_{mk} -\tau_{ml}}{\tau_{kl}}\right).$$ Substituting $\pi_m^*$ into $$\frac{\partial  \tau_{kl}}{\partial E_m}=\frac{\partial  \ln\tau_{kl}}{\partial E_m}\tau_{kl}=-\pi_m^*\tau_{kl}$$ gives the desired result.

Further, using the Kemeny's constant and $\sum_m \pi_m =1$, it is straightforward to show that 
\begin{equation}
    \sum_m \frac{\partial  \tau_{kl}}{\partial E_m} = -\sum_m\pi_m \tau_{kl} +\sum_{m}\pi_m(\tau_{mk}-\tau_{ml})=-\tau_{kl},
\end{equation}
which is exactly the summation response relation
\begin{equation}
    \sum_m \frac{\partial  \ln\tau_{kl}}{\partial E_m} =-1
\end{equation}
in the main text.

\paragraph{Energy-barrier response relation}
We next prove the relation
\begin{equation}\label{SRRmfpt}
\sum_{m<n}\frac{\partial \ln \tau_{kl}}{\partial B_{mn}}=1    
\end{equation}
in the main text. The proof is similar to the proof for energy perturbation:
\begin{align}
    &\sum_{m<n}\frac{\partial\tau_{kl}}{\partial B_{mn}}\nonumber\\&=-\sum_{m\neq n}\frac{\partial  \tau_{kl}}{\partial \ln W_{mn}}=\sum_{m,n} W_{mn} \pi_n (\tau_{kn}-\tau_{km})(\tau_{kl}+\tau_{nk}-\tau_{nl}) \\
    &=\sum_{n\neq k} \pi_n\left[\sum_{m (\neq n)} W_{mn}(\tau_{kn}-\tau_{km})\right](\tau_{kl}+\tau_{nk}-\tau_{nl}) + 0 \label{trick}\\
    &=\sum_{n\neq k}\pi_n \cdot 1\cdot \tau_{kl} +[0-\pi_k(-\tau_{kl})]=\tau_{kl}. \square
\end{align}
It should be noted that in Eq. \eqref{trick}, we divide the terms in the summation into $n=k$ and $n\neq k$, because $\sum_{m}W_{mn}(\tau_{kn} -\tau_{km})=1$ only holds for $m\neq k$. The $m=k$ term vanishes thanks to the zero factor $(\tau_{kl}+\tau_{kk}-\tau_{kl})=0$ so it does not contribute to the sum.
Using this idea we can further prove a relation for the response of steady-state distributions:
\begin{equation}
    \sum_{e_{mn}}\frac{\partial\pi_k}{\partial B_{mn}}=\sum_{m,n}\frac{\partial \pi_k}{\partial W_{mn}}W_{mn}=0,
\end{equation}
which is a key result in \cite{aslyamov2024General}. The derivation is given as:
\begin{align}
    \sum_{m,n}\frac{\partial \pi_k}{\partial W_{mn}}W_{mn}&=\sum_{m,n}W_{mn}\pi_n (\tau_{kn}-\tau_{km})\pi_k \\
    &= \pi_k \sum_{n \neq k}\pi_n \sum_m W_{mn}(\tau_{kn}-\tau_{km})-\pi_k^2\sum_m W_{mk}\tau_{km} \\
    &=\pi_k\sum_{n \neq k}\pi_n - \pi_k^2\sum_m W_{mk}\tau_{km} \\
    &=\pi_k (1-\pi_k)-\pi_k^2\left(\frac{1}{\pi_k}-1\right)=0.
\end{align}
The second last equality is due to $\sum_m W_{mk}\tau_{km}+W_{kk}T_{kk}=-1$ by the recursion relation \eqref{return}, where $T_{kk}=1/(-W_{kk}\pi_k)$ is the mean first return time of state $k$.

\textit{Physical intuition of the summation relation \eqref{SRRmfpt}---uniform global perturbation on the timescale}
This relation can be intuitively understood as follows. It is equivalent to rescale all transition rates by a single multiplicative factor $\theta$ ($B_{ij}\to B_{ij}+\delta B$ for all $i,j$)

\begin{equation}
    W_{ij}\to W_{ij} \theta,
\end{equation}
with $\theta = e^{-\delta B}$. Consequently, an arbitrary MFPT is rescaled by a factor $\theta^{-1}=e^{\delta B}$. Therefore,
\begin{align}
    &\sum_{m<n}\frac{\partial \tau_{kl}}{\partial B_{mn}}=\sum_{m,n}\frac{\partial \tau_{kl}}{\partial \ln W_{mn}}=\frac{\partial \tau_{kl}}{\partial B}=\lim_{\delta B \to 0}\frac{\tau_{kl} e^{\delta B}-\tau_{kl}}{\delta B}=\tau_{kl}\\
    &\Rightarrow \sum_{m<n}\frac{\partial \ln\tau_{kl}}{\partial B_{mn}}=1.
\end{align}
This rescaling of timescale certainly will not change the steady-state distribution, so we also have $\sum_{e_{mn}}\frac{\partial\pi_k}{\partial B_{mn}}=0$.

\paragraph{Nonlinear response relation to strong Energy perturbation}
For a strong perturbation $E_m\to E_m +\Delta E_m$, we show in the main text that:
\begin{equation}
    \frac{\tau_{kl}^{\prime}-\tau_{kl}}{1-e^{-\Delta E_m}}=\frac{\partial  \tau_{kl}}{\partial E_m},
\end{equation}
which is proven using the relations
\begin{align}
    &\pi_m^{\prime}=\frac{e^{-\Delta E_m}\pi_m}{(e^{-\Delta E_m}-1)\pi_m+1}\\
    &\pi_k^{\prime}=\frac{\pi_k}{(e^{-\Delta E_m}-1)\pi_m+1}.\quad(k\neq m)
\end{align}
derived in \cite{bao2024nonequilibrium}. Using the above equation, we get
\begin{align}
    &\frac{\tau_{kl}^{\prime}}{\tau_{kl}}=\frac{\pi_k^*}{\pi_k^{* \prime}}=(e^{-\Delta E_m}-1)\pi_m^*+1 \\
    & \Rightarrow \tau_{kl}^{\prime}-\tau_{kl}=(e^{-\Delta E_m}-1)\tau_{kl}\pi_m^*=(1-e^{-\Delta E_m})\frac{\partial  \tau_{kl}}{\partial E_m}. \square
\end{align}

We can also obtain a similar relation for edge perturbation as:
\begin{equation}
    \frac{\tau_{kl}^{\prime}-\tau_{kl}}{\Delta W_{mn}}=\frac{\tau_{kl}^{\prime}\frac{\partial \tau_{kl}}{\partial W_{mn}}}{\tau_{kl}\Bigl(1+\tau_{nm}\,\pi_n\,\Delta W_{mn}\Bigr)
        -\pi_n\,\Delta W_{mn}\Bigl(\tau_{kn}-\tau_{km}\Bigr)
        \Bigl(\tau_{nk}-\tau_{nl}\Bigr)},
\end{equation}
which has a more compelling form for the specific case of local perturbation ($k=m,\ l=n$):
\begin{equation}
    \frac{\tau_{mn}^{\prime}-\tau_{mn}}{\Delta W_{mn}}=\frac{\tau_{mn} ^{\prime}}{\tau_{mn}}\frac{\partial \tau_{mn}}{\partial W_{mn}}.
\end{equation}

\subsection{Proof of fundamental limits on relative log-response}

By the non-negativity of $\tau'_{kl}$ (or equivalently, $\pi^{* \prime}_n$), we can obtain the two novel inequalities for MFPTs in the original network, as reported in the main text. 
To proceed, recall that the simplified closed-form expression of $\tau'_{kl}$ is given by Eq.~\eqref{update_MPFT}. 

Notably, the denominator of $\tau'_{kl}$ is non-negative for physical perturbations satisfying $\Delta W_{mn}\geq -W_{mn}$ (as required in the main text): 
\begin{equation}
    1+\pi_n\Delta W_{mn}(\tau_{nm}+\tau_{kn}-\tau_{km})\geq 1-W_{mn}\pi_n(\tau_{nm}+\tau_{mn})\geq 0,
\end{equation}
where we use the triangle inequality and the inequality $W_{mn}\pi_n(\tau_{nm}+\tau_{mn})\leq 1$ derived in \cite{bao2024nonequilibrium}. Therefore, the sign of $\tau'_{kl}$ is solely determined by its numerator, which should be non-negative:
\begin{equation}
    \tau_{kl}(1+\tau_{nm}\pi_n\Delta W_{mn})-\pi_n\Delta W_{mn}(\tau_{kn}-\tau_{km})(\tau_{nk}-\tau_{nl})\geq 0.
\end{equation}
Taking the $\Delta W_{mn}\to\infty$ and $\Delta W_{mn}=-W_{mn}$ limits on the above equation yields the two desired inequalities:
\begin{align}
    & (\tau_{kn}-\tau_{km})
       (\tau_{nk}-\tau_{nl})\leq\tau_{kl}\tau_{nm},\\
      & \phi_{mn}\left(\tau_{nm}-\frac{(\tau_{kn}-\tau_{km})
       (\tau_{nk}-\tau_{nl})}{\tau_{kl}}\right)\leq 1, \label{bound2}
\end{align}
which has been presented in Eq. (7) of the main text. The log-response reads
\begin{equation*}
    \frac{\partial\ln\tau_{kl}}{\partial \ln W_{mn}}=-\phi_{mn} (\tau_{kn}-\tau_{km})\left(1+\frac{\tau_{nk}-\tau_{nl}}{\tau_{kl}}\right)
\end{equation*}
Combining the two inequalities and the triangle inequality for MFPTs with the above expression, we obtain the desired lower bound 
(we only need to consider $\tau_{kn}\geq\tau_{km}$ to obtain the lower bound, where the sign of the response is negative)
\begin{equation}\label{local3}
    \frac{\partial\ln\tau_{kl}}{\partial \ln W_{mn}}\geq \frac{\partial\ln\tau_{mn}}{\partial \ln W_{mn}}=-\phi_{mn}(\tau_{mn}+\tau_{nm})\geq -1
\end{equation}
and upper bounded as (we only need to consider $\tau_{kn}\leq\tau_{km}$ to obtain the upper bound)
\begin{equation}
    \frac{\partial\ln\tau_{kl}}{\partial \ln W_{mn}}\leq \phi_{mn}\left(\tau_{nm}-\frac{(\tau_{kn}-\tau_{km})
       (\tau_{nk}-\tau_{nl})}{\tau_{kl}}\right) \leq 1.
\end{equation}

\section{Additional numerical results for the folding example}
{The folding network of Fig.~1 of the main text uses, in units of the productive folding rate $W_{NI}=1$, the rates $W_{IU}=1$, $W_{UI}=0.3$, $W_{IN}=0.01$, $W_{MI}=0.2$ and $W_{IM}=0.05$, and, when the chaperone is present, $W_{CM}=0.2$ and $W_{NC}=W_{IC}=1$, so that half of the chaperone cycles are productive. The same network is used for the panels below.} Panel~(a) shows the upstream accessibility $U_{N\leftarrow l\mid I}$ for different initial states $l$ in the chaperone network. Because $G_{N\mid MI}$ and $\Sigma_{N\mid MI}$ are fixed there, the change in response amplitude comes entirely from how easily trajectories from $l$ reach the perturbed source state $I$ before hitting $N$. Panel~(b) shows the downstream gain/loss $G_{N\mid MI}=\tau_{NI}-\tau_{NM}$ as a function of the chaperone capture rate $W_{CM}$. Once $M$ becomes kinetically closer to $N$ than $I$ is, increasing $W_{MI}$ turns from harmful to helpful.
\begin{figure}[!htb]
    \centering
    \includegraphics[width=.8\textwidth]{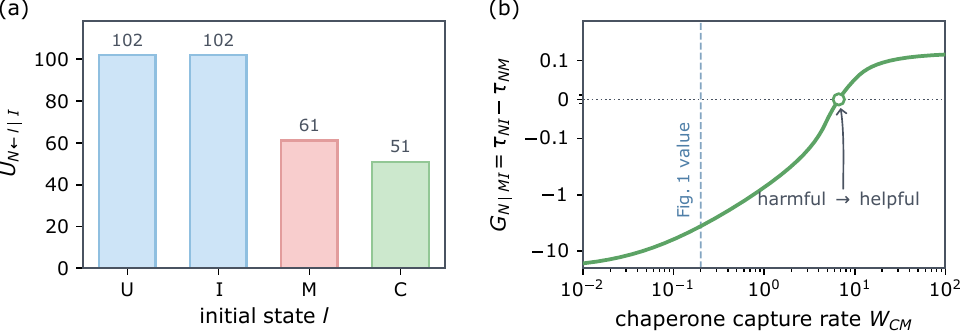}
    \caption{{Additional numerical panels for the folding example. Panel~(a) gives the upstream accessibility in the chaperone network for different initial states. Panel~(b) gives the downstream gain/loss as the chaperone capture rate $W_{CM}$ is varied {; it changes sign at $W_{CM}\simeq6.7$, well above the value used in Fig.~1 (dashed line)}.}}
    \label{fig:sm_cd}
\end{figure}

\bibliographystyle{apsrev4-2}
\bibliography{ref}